\documentclass[twocolumn,prb,showpacs]{revtex4}

\usepackage{graphicx}
\usepackage{amsmath}
\usepackage{amsfonts}
\usepackage{amssymb}
\usepackage{enumerate}
\usepackage{hyperref}
\hypersetup{%
  breaklinks = {true}, 
  colorlinks = {true}, 
  citecolor = {blue},  
  filecolor = {black}, 
  linkcolor = {red},   
  urlcolor  = {blue},  
  pdfauthor = {\textcopyright\ Carsten Raas and G\"otz S. Uhrig},
  pdfstartpage = {1},
  pdfstartview = {FitH}, 
}

\begin{document}

\title{Generic susceptibilities of the half-filled Hubbard model in infinite dimensions}

\author{Carsten Raas}
\email{carsten.raas@tu-dortmund.de}
\homepage{http://www.raas.de}
\affiliation{Lehrstuhl f\"{u}r Theoretische Physik I,
  Technische Universit\"{a}t Dortmund, Otto-Hahn Stra\ss{}e 4, 44221 Dortmund, Germany}

\author{G\"otz S. Uhrig}
\email{goetz.uhrig@tu-dortmund.de}
\homepage{http://t1.physik.tu-dortmund.de/uhrig/}
\affiliation{School of Physics, University of New South Wales,
  Kensington 2052, New South Wales, Australia}
 \altaffiliation{On leave from Lehrstuhl f\"{u}r Theoretische Physik I,
    Technische Universit\"{a}t Dortmund, Otto-Hahn Stra\ss{}e 4,
    44221 Dortmund, Germany}

\date{\textrm{\today}}

\begin{abstract}
  Around a metal-to-insulator transition driven by repulsive
  interaction (Mott transition) the single particle
  excitations and the collective excitations are equally important.
  Here we present results for the generic susceptibilities at zero
  temperature in the half-filled Hubbard model in infinite
  dimensions. Profiting from the high resolution of dynamic
  density-matrix renormalization at all energies, results for the
  charge, spin and Cooper-pair susceptibilities in the metallic and
  the insulating phase are computed. In the insulating phase, an
  almost saturated local magnetic moment appears. In the metallic
  phase a pronounced low-energy peak is found in the spin response.
  It is the precursor of the magnetic moment in the insulator.
\end{abstract}


  
  

\maketitle


\section{Introduction}
\label{sec:intro}

Strongly correlated systems persist to be a very interesting field
of current research. In particular in the vicinity of a quantum
phase transition the physics is very rich because the nature of the
ground state and of the excitations changes. One prominent example
of such a phase transition at zero temperature is the
metal-to-insulator transition driven by an increasing local
repulsive interaction: the so-called Mott transition. For low values
of the interaction the system is metallic because the electrons can
still pass one another. For large values of the repulsive Coulomb
interaction, there can be at most one electron per site. If the
system is exactly half-filled, each site is occupied by one
electron. No motion of electrons is possible because they are
blocking one another. Hence the system is insulating.

The simplest model describing the Mott transition is the Hubbard
model. \cite{hubba63,kanam63,gutzw63} The one-dimensional case can
be solved analytically and has been studied
intensively.\cite{essle05} It is governed by the particular
phenomena of one-dimensional physics such as spin-charge
separation. Yet this is not the generic physics occurring in higher
dimensions. Very much of our current understanding of the Mott
transition in higher dimensions is based on the limit of infinite
dimensions\cite{metzn89a,mulle89a} which leads to the dynamic
mean-field theory (DMFT) (Refs.\ \onlinecite{prusc95} and
\onlinecite{georg96}) as important approximation scheme for real
narrow-band compounds.

\begin{figure}
  \begin{center}
    \includegraphics[width=0.99\columnwidth]{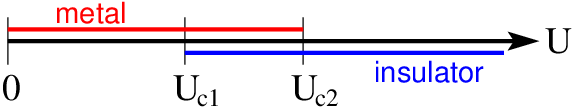}
  \end{center}
  \caption{(color online) Phase diagram of the Mott transition of a
    half-filled Hubbard model in infinite dimensions with
    semi-elliptic density-of-states at zero temperature as function
    of the interaction $U$. The critical value below which the
    insulator ceases to exist is $U_{c1}=2.38\pm 0.02D$. The
    critical value above which the metal ceases to exist is
    $U_{c2}=3.07\pm 0.1D$ (Refs.\ \onlinecite{karsk05} and
    \onlinecite{karsk08}). But between $U_{c1}$ and $U_{c2}$ the
    metal has the lower energy so that the insulator is only
    metastable (Ref.\ \onlinecite{karsk05}).}
  \label{fig:phasdia}
\end{figure}

The essential result of DMFT is that the transition between metal
and paramagnetic insulator is marginally first
order.\cite{georg96,bulla99a,bulla01a,blume02,potth03b,karsk05} It
is first order at finite temperature where a finite amount of
spectral weight is redistributed at the transition $U_c$ with
$U_{c1}(T)<U_c(T)<U_{c2}(T)$. But at zero temperature only an
infinitesimal amount of spectral weight is redistributed at
$U_c(T=0)=U_{c2}(T=0)$ (Ref.\ \onlinecite{karsk05}) so that the
transition is continuous. The first order jump has just vanished.
The insulator represents a metastable phase for $U < U_{c2}$ (Refs.\
\onlinecite{blume02} and \onlinecite{karsk05}) (see Fig.\
\ref{fig:phasdia}).

Historically, there are two scenarios for the Mott transition based
on opposite limits. The Brinkmann-Rice scenario captures the
essential point on the metallic side, namely the
band-narrowing.\cite{brink70} The Hubbard scenario captures the two
bands in the insulator (Hubbard bands), which approach each other
till they touch at the point where the insulator becomes
unstable.\cite{hubba64b} The DMFT combines the strong points of both
preceding
scenarios.\cite{zhang93,georg96,bulla99a,kotli99,bulla01a,karsk05,karsk08}
Already the metallic solution displays Hubbard bands. The
difference between the metallic and the insulating solution is found
in the re-distribution of spectral weight at moderate energies only
while the spectral densities at higher energies coincide. This
coincidence is quantitative for $U\to U_{c2}$ as shown in Ref.\
\onlinecite{karsk05}.

So the generic single-particle dynamics as encoded in the
single-particle propagator is by now well-understood for the Mott
transition. The dynamics of collective excitations and in particular
the interplay between the single-particle modes and the collective
modes is less well understood and investigations are
ongoing.\cite{byczu07a,raas09a}

A truly open issue is the physical origin of sharp features found at
the inner band edges of the Hubbard bands just before the system
switches from metallic to insulating behavior. These features were
observed in quite a number of investigations but they were discussed
as physical phenomenon only recently\cite{karsk05,karsk08} based on
high-resolution dynamic density-matrix renormalization (D-DMRG).
The features are confirmed by independent D-DMRG
calculations\cite{nishi06} and by high-resolution numerical
renormalization (NRG) (Ref.\ \onlinecite{zitko09}) but called into
question by quantum Monte-Carlo extrapolations.\cite{blume08}

For the above reasons we have performed a thorough investigation of
the susceptibilities in the half-filled Hubbard model in infinite
dimensions. The calculations start from the self-consistent
solutions obtained by iterating the DMFT self-consistency cycle
\cite{georg92a,jarre92,georg96} with D-DMRG as impurity
solver.\cite{karsk05,karsk08} We benefit again from the good control
of the energy resolution achievable by D-DMRG for all energies.

The susceptibilities provide valuable complementary information to
the single-particle propagator. They address bosonic observables
such as the local charge or spin or Cooper-pair density and their
dynamics. So they give information about the corresponding
collective modes. Moreover, the susceptibilities are experimentally
relevant. The charge susceptibility corresponds to the
polarizability which determines the response seen in linear optics
such as in infrared absorption. The spin susceptibility can be
measured by inelastic neutron scattering. For these reasons, the
susceptibilities $\chi^\text{charge}$, $\chi^\text{spin}$, and
$\chi^\text{pair}$ are addressed in the present article.

The paper is organized as follows. In Sec.\ \ref{sec:model}, we
introduce the model. In Sec.\ \ref{sec:results}, the
susceptibilities will be defined in detail. Analytic statements
about them will be derived there. The numerical results in the
insulating and in the metallic phases will be given and their
physical implications will be discussed. The conclusions summarize
the article.


\section{Model and Method}
\label{sec:model}

We study the Hubbard model in DMFT at half-filling. The Hamiltonian
reads
\begin{equation}
  \label{eq:hamilton}
  \mathcal{H} = -t \sum_{\langle i, j\rangle; \sigma}
  c^\dagger_{i;\sigma}
  c^{\phantom\dagger}_{j;\sigma} + U \sum_i (\hat n_{i;\uparrow}-1/2)
  (\hat n_{i;\downarrow}-1/2).
\end{equation}
Here $c^\dagger_{i;\sigma}$ creates a fermion of spin $\sigma$ at
site $i$ while $c_{i;\sigma}$ annihilates such a fermion. The matrix
element $t$ labels the hopping of the fermions from site to site,
i.e., their kinetic energy. The matrix element $U>0$ labels the
on-site Coulomb repulsion of two electrons on the same site. It is
an effective parameter which takes the screening into account.

We assume that the lattice on which it is defined is bipartite so
that the Hamiltonian \eqref{eq:hamilton} displays particle-hole
symmetry. In our physical discussion we will assume that the system
is translationally invariant so that the momentum dependence of a
propagator can be considered. The actual calculations, however, will
be done for a non-interacting semi-elliptic density-of-states (DOS)
\begin{equation}
  \rho_0(\omega) = (2/(\pi D^2))\sqrt{D^2-\omega^2}
\end{equation}
which is characteristic for the Bethe lattice with infinite
branching ratio.\cite{econo79} The advantage of this approach is
calculational simplicity and a finite support in contrast to the
Gaussian tails on truly hypercubic lattices.\cite{metzn89a,mulle89a}

The limit of infinite dimensions $d$ or equivalently of an infinite
coordination number $z$ exists if $t$ is scaled like
$t^\star/\sqrt{z}$. This limit defines the
DMFT.\cite{metzn89a,mulle89a,georg92a,jarre92,georg96} The
self-energy $\Sigma_{ij}$ becomes local
$\Sigma_{ij}=\delta_{ij}\Sigma_{ii}$ and equals the self-energy of a
single-impurity Anderson model\cite{hewso93} which has the same
skeleton diagrams in all orders. This means that the local dressed
propagator $G_{ii}$ must be the same which defines the
self-consistency condition of the DMFT. The ensuing simplification
is that one only has to solve a zero-dimensional problem: an
interacting site coupled to a bath. One way to represent this bath
is as semi-infinite chain so that the problem is amenable to
powerful one-dimensional tools, for instance dynamic
DMRG\cite{garci04,karsk05} which ensures a good control of the
energy resolution over all energies.\cite{raas04a} For details, we
refer the reader to Ref.\ \onlinecite{karsk08}.


\section{Susceptibilities}
\label{sec:results}

In a translationally invariant system it is appropriate to consider
the momentum-dependent $\chi(\mathbf{q})$. But the implications of
the limit of infinite dimensions are easier seen in real space. In
real space $\chi$ depends only on the difference
$\mathbf{r}_i-\mathbf{r}_j$ between the site $i$ where the
observable is measured and the site $j$ where the field is
applied. In the limit $d\to\infty$ each fermionic propagator from
$i$ to $j$ is scaled by a factor $d^{-||i-j||/2}$ where $||\cdot||$
is the taxi cab or New York metric which counts the minimum number
of hops required to get from $i$ to $j$. In a diagrammatic
description of the propagation of any bosonic, collective observable
from $i$ to $j$ at least two fermionic propagators link $i$ and
$j$. This implies that such a susceptibility is suppressed by
$d^{-||i-j||}$. Hence only \emph{local} susceptibilities from $i$ to
$i$ matter in infinite dimensions, at least in the absence of phase
transitions.

This conclusion is not quite the whole story because certain
non-local contributions can add up. For instance there are $2d$
next-neighbor contributions ($||i-j||=1$) to $i$ so that they make a
non-negligible contribution if they add up since $2d/d=2$. But they
will not add up for a generic momentum $\mathbf{q}$. The momentum
$\mathbf{q}$ enters the susceptibilities only via
\begin{equation}
  \eta(\mathbf{q}) = \frac{1}{d}\sum_{p=1}^d\cos(q_p)
\end{equation}
where $q_p$ is the component in direction
$p$.\cite{mulle89a,uhrig93b} In the limit $d\to\infty$ almost all
vectors $\mathbf{q}$ imply $\eta(\mathbf{q})=0$ since
$\sqrt{d}\eta(\mathbf{q})$ is gaussian distributed with finite
variance. Only particular values of measure zero, for instance
$\mathbf{q}=(\pi,\pi,\pi,\ldots,\pi)^\dagger$, imply a non-vanishing
$\eta$. Hence, the generic susceptibility is the one for $\eta=0$
which corresponds in real space to the local one.

Of course, some important effects of finite-dimensional physics are
not captured by the generic susceptibilities. For example an
antiferromagnetic instability or the instability to an
incommensurate phase is indicated by the divergence of a
susceptibility for some $\eta \neq 0$. This must be kept in mind.
On the other hand, however, the propagation of collective modes as
they interact with single particles is given in $d=\infty$ by the
generic susceptibilities. In any diagram for the proper self-energy
which contains the propagation of a collective mode there is also a
sum over its momentum. Hence the peculiar contributions with
$\eta\neq0$ do not matter here. It is the generic behavior of
collective modes at $\eta=0$ which is relevant for the interaction
with single particles in $d=\infty$.

Note that this argument remains true even if
the DMFT is not seen as the limit of infinite dimensions but more broadly as
a consistent local approximation scheme. This is by now a very common view
adopted in the description of real compounds. Then one can discuss
the whole momentum dependence of the susceptibility. But the
behavior of the collective modes which enters implicitly in the description
of the single-particle dynamics remains the one given by the local
susceptibilities. For this reason, we compute and discuss
the local susceptibilities in the following.

The local susceptibilities are easily accessible since they are
identical to the local susceptibilities at the interaction site of
the auxiliary single-impurity Anderson model. This is obvious if
one thinks of the susceptibilites as being given by an expansion in
terms of skeleton diagrams to infinite order.

In the sequel, we will present the imaginary parts of the
susceptibilities since these parts provide direct information on the
energies and spectral weights of collective excitations.

One drawback of the susceptibilities is that their imaginary parts
are antisymmetric (odd) by construction $\text{Im}\chi(\omega) =
-\text{Im}\chi(-\omega)$. Hence the very interesting behavior at
zero and at very low energies is suppressed. For instance, let $Q$
be an hermitean bosonic observable like the spin density. With
$|0\rangle$ being the ground state and $E_0$ its energy, the matrix
element of the resolvent
\begin{equation}
  \label{eq:resolv-def}
  R(\omega) := \left\langle0\big|Q
    (\omega-(\mathcal{H}-E_0))^{-1}Q\big|0\right\rangle
\end{equation}
can have a $\delta$ peak at $\omega=0$ in its imaginay part. This
would constitute an important piece of information on the
system. But the susceptibility $\chi$ constructed from $R$ according
to
\begin{equation}
  \label{eq:chi-resolv}
  \chi(\omega) = -R(\omega+i0+)-R(-\omega-i0+)
\end{equation}
would not display this $\delta$ peak because it cancels on the right
hand side of \eqref{eq:chi-resolv}. In order not to lose the
information at zero and at very low frequencies we will display
results for
\begin{equation}
  \label{eq:chi-plus-resolv}
  \text{Im}\chi_+(\omega)=\text{Im}R(\omega)
\end{equation}
which is the contribution for non-negative frequencies. The
contribution for negative frequencies is implied by antisymmetry,
i.e., $\text{Im}\chi_-(\omega) =-\text{Im}\chi_+(-\omega)$.

An important tool in understanding spectral densities are sum
rules. From the relation \eqref{eq:chi-plus-resolv}, the definition
\eqref{eq:resolv-def}, and the Hilbert representation of $R(\omega)$
it is obvious that the total spectral weight takes the value
\begin{subequations}
  \label{eq:sumrule}
  \begin{eqnarray}
    \int_0^\infty \chi_+(\omega)d\omega
    &=& \pi \lim_{\omega\to\infty} \omega R(\omega)\\
    &=& \pi \left\langle0\big|Q^2\big|0\right\rangle.
  \end{eqnarray}
\end{subequations}
So knowing the ground state expectation value of $Q^2$ helps to
understand general trends in spectral weights and it provides an
important check for numerical calculations

A last important point concerns the computation of spectral
densities like $\text{Im}\chi_+(\omega)$ by D-DMRG. The dynamic DMRG
calculates a correction vector $|\text{cv}\rangle$ besides the
ground state $|0\rangle$ and the state obtained from the application
of $Q$, $Q|0\rangle$.\cite{ramas97,kuhne98} This correction vector
reads
\begin{equation}
  \label{eq:corrvec}
  |\text{cv}\rangle = (\omega+i\delta-(H-E_0))^{-1}Q|0\rangle .
\end{equation}
Obviously, $\langle0|Q |\text{cv}\rangle$ yields
$R(\omega+i\delta)$. The computation of the correction vector
requires a numerically demanding matrix inversion.\cite{raas04a} If
there is an inaccuracy $\varepsilon$ in the correction vector
$|\text{cv}\rangle$ the imaginary part of $R(\omega+i\delta)$ can be
obtained from a variational functional with a decreased inaccuracy
of the order of $|\varepsilon|^2$, see Ref.\ \onlinecite{jecke02}.

In any case, the numerical approaches require the imaginary
frequency $\delta$ in \eqref{eq:corrvec} to be finite. This implies
a certain broadening of the actual spectral density. In order to
retrieve the unbroadened spectral density $\text{Im}\chi_+(\omega)$
we employ the non-linear least-bias (LB) deconvolution
technique.\cite{raas05a} This approach yields always non-negative
results as is to be expected for $\text{Im}\chi_+(\omega)$. For
details of the approach we refer the reader to Ref.\
\onlinecite{raas05a}. Any deviations from the procedure described
therein will be given below where applicable.


\subsection{Cooper Pair Susceptibility}

Here we consider the local observable
\begin{equation}
  \label{eq:qpair}
  Q^\text{pair} = c^\dagger_{i;\uparrow}c^\dagger_{i;\downarrow}
  +c^{\phantom\dagger}_{i;\downarrow}c^{\phantom\dagger}_{i;\uparrow}
\end{equation}
which creates or annihilates a Cooper pair on site $i$. Of course,
one would expect that the corresponding susceptibility
$\chi^\text{pair}(\omega)$ is strongly suppressed in a Hubbard band
with repulsive interaction. Yet it can contain interesting features,
for instance at higher energy.

But it is not necessary to compute $\chi^\text{pair}(\omega)$
separately. Indeed, there is an underlying symmetry which links
$Q^\text{pair}$ to $Q^\text{charge}$, see Eq.\ \eqref{eq:qcharge}
below. As a result the pair susceptibility is identical to the
charge susceptibility. So we will not discuss
$\chi^\text{pair}(\omega)$ here but refer to the next subsection
where $\chi^\text{charge}(\omega)$ is investigated.

The symmetry becomes apparent under the transformation
$c^\dagger_\sigma \to \gamma^\dagger_\sigma$ according to
\begin{subequations}
  \label{eq:trafo}
  \begin{eqnarray}
    \gamma^\dagger_\uparrow
    &:=& c^\dagger_\uparrow \cos\varphi - c^{\phantom\dagger}_\downarrow\sin\varphi\\
    \gamma^\dagger_\downarrow
    &:=& c^\dagger_\downarrow \cos\varphi + c^{\phantom\dagger}_\uparrow\sin\varphi.
  \end{eqnarray}
\end{subequations}
On a bipartite lattice this transformation is performed on all even
sites with the angle $\varphi$ and on all odd sites with the angle
$-\varphi$. Then the hopping terms in \eqref{eq:hamilton} are left
invariant, independent of the value of $\varphi$. The same is true
for the onsite interaction as given by the term proportional to $U$
in \eqref{eq:hamilton}. So the total Hamiltonian \eqref{eq:hamilton}
remains invariant under \eqref{eq:trafo}.

The interesting relation is the one for the observables
\begin{equation}
  \label{eq:observ-trafo}
  Q^\text{charge}_c  = Q^\text{charge}_\gamma \cos(2\varphi)
  + Q^\text{pair}_\gamma \sin(2\varphi)
\end{equation}
where we use $Q^\text{charge}$ in anticipation of Eq.\
\eqref{eq:qcharge}. The subscript $_c$ refers to the expression in
terms of the original fermions $c$ and $c^\dagger$ while the
subscript $_\gamma$ refers to the expression in terms of the
transformed fermions $\gamma$ and $\gamma^\dagger$. Here the value
of the angle of rotation $\varphi$ matters. For $\varphi=\pi/4$ we
switch from the charge observable to the pair observable. Hence the
corresponding local susceptibilities are indeed the same. No
additional numerics is needed in the bipartite half-filled case.

The above symmetry transformation has not gone unnoticed. It is one
of the transformations at the basis of the SO(5) theory which is
comprehensively reviewed in Ref.\ \onlinecite{demle04}. One further
important conclusion is that charge and superconducting order are
degenerate as far as they stem from a bipartite Hubbard model at
half-filling, for instance with negative $U$.


\subsection{Charge Susceptibility}

Here we consider the local observable
\begin{equation}
  \label{eq:qcharge}
  Q^\text{charge} = c^\dagger_{i;\uparrow}c^{\phantom\dagger}_{i;\uparrow}
  +c^\dagger_{i;\downarrow}c^{\phantom\dagger}_{i;\downarrow} -1
\end{equation}
which measures the charge fluctuations around half-filling, i.e.,
the deviation of the total fermion number per site from $1$.

The general sum rule \eqref{eq:sumrule} requires the expectation
value $\left\langle0\big|Q^2\big|0\right\rangle$ which amounts for
$Q^\text{charge} $ up to twice the double occupancy value. First we
note
\begin{equation}
  (Q^\text{charge})^2 = 2 \hat n_\uparrow \hat n_\downarrow
  - (n_\uparrow +\hat n_\downarrow) +1 
\end{equation}
where $\hat n_\sigma = c^\dagger_{i;\sigma}c^{\phantom\dagger}_{i;\sigma}$. 
This implies at half-filling
\begin{equation}
  \label{eq:charge-sumrule}
  \int_0^\infty \chi_+^\text{charge}(\omega)d\omega
  = 2 \pi \left\langle0\big|\hat n_\uparrow \hat n_\downarrow\big|0\right\rangle.
\end{equation}
So the static quantity to be known for the sum rule is the double
occupancy (see Refs.\ \onlinecite{nishi04b} and
\onlinecite{karsk05}).


\subsubsection{Insulator}

\begin{figure}
  \begin{center}
    \includegraphics[width=0.99\columnwidth]{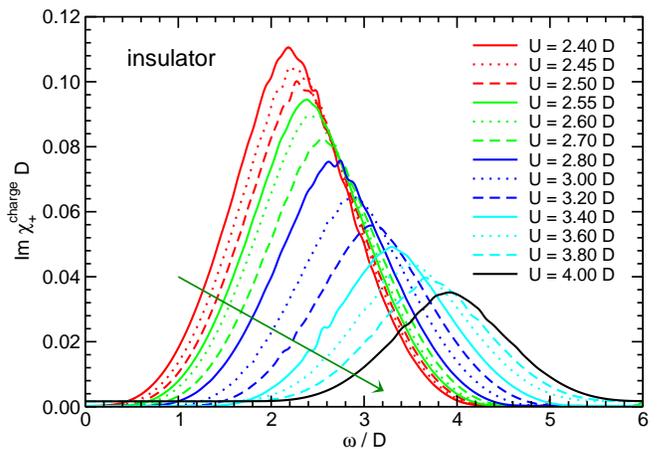}
  \end{center}
  \caption{(color online) Positive imaginary part
    $\chi^\text{charge}_+$ of the local charge susceptibility in the
    insulating phase as function of frequency for various values of
    the interaction $U$. The arrow points in the direction of
    increasing interaction.}
  \label{fig:charge-insulator}
\end{figure}

In Fig.\ \ref{fig:charge-insulator} a series of numerical results is
shown for the positive imaginary part $\chi^\text{charge}_+$ of the
local charge susceptibility in the insulating phase. In the D-DMRG
we kept $m=128$ states in the truncated DMRG basis. The mesh of
frequencies is given by the interval $\Delta\omega=0.05D$ and the
imaginary broadening was $\delta=0.1D$. The LB
deconvolution\cite{raas05a} was perforemd with a tolerance constant
of $1/A_\text{LB}=1/100$. The curves are not perfectly smooth but
display some wiggles. This is due to the deconvolution procedure
employed.

Physically, no special features are discernible. But two trends are
clearly visible. First, the susceptibility is more and more
suppressed as the interaction is increased. This results in an
overall reduction of the area under the curves. It can be quantified
by the sum rule \eqref{eq:charge-sumrule}. So it is natural that the
spectral weight of the charge response decreases on increasing $U$
because the latter suppresses the double occupancy $\langle0|\hat
n_\uparrow \hat n_\downarrow|0\rangle$ more and
more.\cite{nishi04b,karsk05} We have checked this sum rule
numerically and found it to be fulfilled to within a relative error
of $1.8 \%$ on the deconvolved data in the interval $\omega\in [0,
5.9D]$.

Second, the spectral weight is shifted to higher and higher
frequencies on increasing interaction. This is seen in two
features. One is the peak position which is shifted. Its shift
corresponds to the shift of spectral weight in the single-particle
propagators.\cite{nishi04b,karsk05,karsk08} These shifts reflect the
simple fact that the energy difference between the lower and the
upper Hubbard band is given by about $U$. Hence it increases
linearly with $U$.

The other feature is the onset of finite spectral density which
increases also with $U$. Due to the LB deconvolution\cite{raas05a}
there is no region where the spectral density is strictly zero. But
we have checked that the susceptibility data is perfectly consistent
with the natural assumption that the onset of the
$\text{Im}\chi^\text{charge}_+$ takes place at $2\Delta$ where
$\Delta$ is the single-particle gap (for data, see
\onlinecite{karsk05} and \onlinecite{karsk08}). For this check we
analyzed the susceptibility data by fitting a quadratic onset
$\propto (\omega-\omega_\text{onset})^2$ plus higher order
corrections to its continuum. The quadratic onset is to be expected
from the square root onset of the single-particle
bands\cite{nishi04b,karsk05} which is convolved with itself in the
standard particle-hole bubble. Because the single-particle gap rises
upon increasing $U$ the collective response is shifted to higher and
higher energies.

The onset at $2\Delta$ reflects the fact that the collective mode is
made from a particle and a hole. In the particle-hole symmetric case
considered here both have the same gap $\Delta$ so that any
collective continuum starts only at $2\Delta$. A lower onset, i.e.,
at lower energy, could arise only if binding transferred spectral
weight to the frequency interval below $2\Delta$. No indication for
such a binding is found here.


\subsubsection{Metal}

\begin{figure}
  \begin{center}
    \includegraphics[width=0.99\columnwidth]{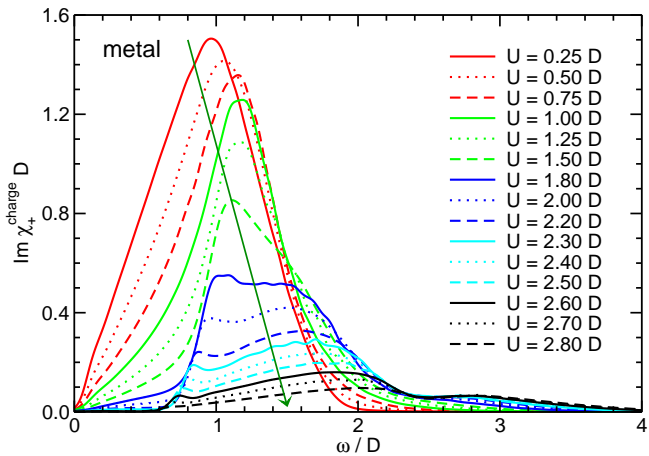}
  \end{center}
  \caption{(color online) Positive imaginary part
    $\chi^\text{charge}_+$ of the local charge susceptibility as
    function of frequency for various values of the interaction $U$
    in the metallic phase. The arrow points in the direction of
    increasing interaction. Note the different scale of the response
    compared to the response in the insulator in Fig.\
    \ref{fig:charge-insulator}.}
  \label{fig:charge-metal-all}
\end{figure}

In Fig.\ \ref{fig:charge-metal-all} a series of numerical results is
shown for the positive imaginary part $\chi^\text{charge}_+$ of the
local charge susceptibility in the metallic phase. In the D-DMRG we
kept between $m=128$ to $m=256$ states in the truncated DMRG basis;
the frequency mesh is given by $\Delta\omega=0.05D$ and the
imaginary broadening by $\delta=0.1D$.  The LB
deconvolution\cite{raas05a} was performed with tolerance constants
$1/A_\text{LB}=1/10$ and $1/A_\text{LB}=1/100$. The curves are not
perfectly smooth but display some minor wiggles. This is due to the
deconvolution procedure employed.

As in the insulating regime we find the trend that increasing
interaction suppresses the charge response. This is expected because
it is related to the same sum rule \eqref{eq:charge-sumrule} which
holds independent of the phase under study. Furthermore, the
spectral response is shifted to higher and higher energy. Again this
general trend can be related to the same trend in the
single-particle propagators.\cite{karsk05,karsk08}

Interestingly, the charge response in the metallic phase displays
much more structure than in the insulating phase. For low values of
$U$ we find a linear increase with frequency $\omega$ for not too
high values of $\omega\lessapprox 0.6D$. This is the expected
behavior for a Fermi liquid. Its slope becomes smaller and smaller
as the quasiparticles become heavier and heavier. From $U\approx
1.5D$ onwards, most of the charge response lies in an intermediate
range $D\lessapprox \omega \lessapprox 2D$. There is still some
spectral weight at lower frequencies but it is decreasing rapidly.
An additional shoulder situated between $0.6D$ and $1D$ occurs above
$U=1.8D$. We will discuss this feature in detail below. Above
$U\approx 2.2D$ there is a third rather flat hump discernible
centered around $\omega=3D$.

Qualitatively, the three regions of charge response can be
understood from the single-particle response. The single-particle
response, see for instance Figs.\ 12 and 13 in Ref.\
\onlinecite{karsk08}, is mainly characterized by the heavy
quasiparticle in the narrow central peak and by the broad emerging
Hubbard bands of significant weight which are centered around
$\omega \approx 1.5D$. Assuming that the collective response is
roughly given by a single diagrammatic particle-hole bubble (or by
an analytic function of this bubble as in the random phase
approximation or in more sophisticated approaches such as the local
moment approach),\cite{logan98,galpi08} we simply have to convolve
the single-particle response at positive frequencies with the one at
negative frequencies.

The response at low frequencies $\omega \lessapprox 0.8D$ stems from
the convolution of the central peak of heavy quasiparticles with
itself. It dominates at low values of $U$, but on increasing $U$ it
decreases in weight like $Z^2$ where $Z$ is the quasiparticle weight
vanishing linearly for $U\to U_{c2}$. The quasiparticle weight $Z$
measures the spectral weight in the central peak.

The response at intermediate frequencies $ 0.8D \lessapprox\omega
\lessapprox 2.2D$ stems from the convolution of the central peak of
heavy quasiparticles with one of the Hubbard bands. Hence it is
higher in frequency, because the Hubbard band is, and its weight
decreases only linearly in $Z$

The response at higher frequencies results from the convolution of
the lower and the upper Hubbard band. Hence it is located around
twice their energy, i.e., around $3D$. This contribution is not
suppressed by $Z$ so that it survives in the limit $U\to
U_{c2}$. This is consistent with our data for $U=2.8D$ shown in
Fig.\ \ref{fig:charge-metal-all}. Unfortunately, no reliable data
even closer to the critical value $U_{c2}$ could be obtained.

\begin{figure}
  \begin{center}
    \includegraphics[width=0.99\columnwidth]{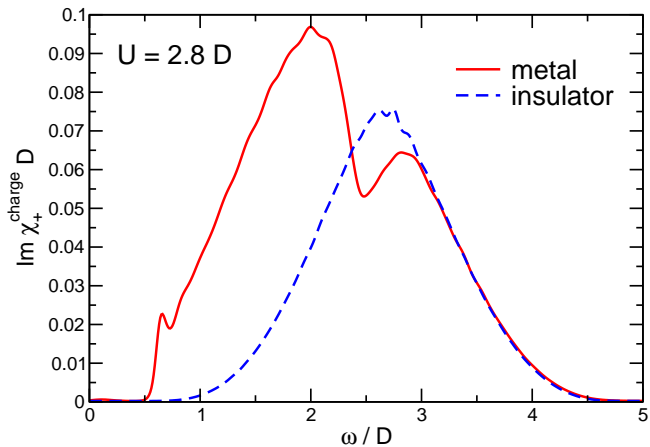}
  \end{center}
  \caption{(color online) Positive imaginary part
    $\chi^\text{charge}_+$ of the local charge susceptibility as
    function of frequency for interaction $U=2.8D$ in the metallic
    and insulating phase.}
  \label{fig:charge-compare}
\end{figure}

Note that only the last contribution at the higher frequency has an
analog in the response in the insulator since there only the Hubbard
bands exist. Indeed, the insulating response describes the high
frequency metallic response very well as is illustrated in Fig.\
\ref{fig:charge-compare}. The insulating and the metallic curve
agree very well above $\omega\approx 3D$. The sizable differences
below this frequency are remarkable in view of the shift of fairly
little spectral weight between the metallic and the insulating
single-particle solution.\cite{karsk05} For experiment, for
instance infrared absorption, Fig.\ \ref{fig:charge-compare}
provides valuable information how different a metallic and an
insulating system can look even though only tiny parameter changes
are made. In the present case, even no parameters are changed, but
only different hysteresis branches are considered.

\begin{figure}
  \begin{center}
    \includegraphics[width=0.99\columnwidth]{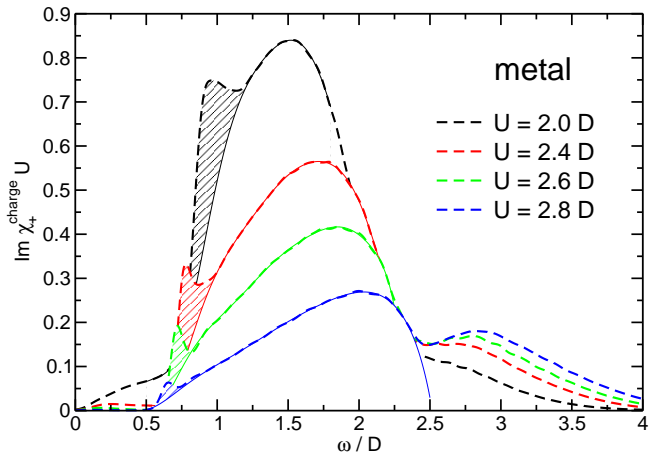}
  \end{center}
  \caption{(color online) Positive imaginary part
    $\chi^\text{charge}_+$ of the local charge susceptibility as
    function of frequency for values of the interaction $U$ in the
    metallic phase, but close to the transition to the
    insulator. The main contribution is extrapolated as if there
    were no additional shoulder (thin solid lines). Then the shaded
    area is attributed to spectral weight of the shoulder.}
  \label{fig:charge-metal-shoulder}
\end{figure}

Let us come back to the shoulder seen between $\omega\approx 0.6D$
and $\omega\approx 1D$ for $U\gtrapprox 1.8D$. Its position
corresponds very precisely to the frequency where the sharp feature
at the inner band edges has been found, see Fig.\ 2 in Ref.\
\onlinecite{karsk05} and Figs.\ 12 and 13 in Ref.\
\onlinecite{karsk08}. Hence it is to be expected that there is a
relation between both features. In view of the hypothesis that the
sharp feature is caused by a resonance made from a heavy
quasiparticle and a collective mode (see Refs.\ \onlinecite{karsk05}
and \onlinecite{karsk08}), it would be appealing to interprete the
shoulder in Fig.\ \ref{fig:charge-metal-all} as the cause for the
sharp feature in the single-particle spectral density. To pursue
this idea further we plot in Fig.\ \ref{fig:charge-metal-shoulder}
the metallic charge response close to the transition to the
insulator.

\begin{figure}
  \begin{center}
    \includegraphics[width=0.99\columnwidth]{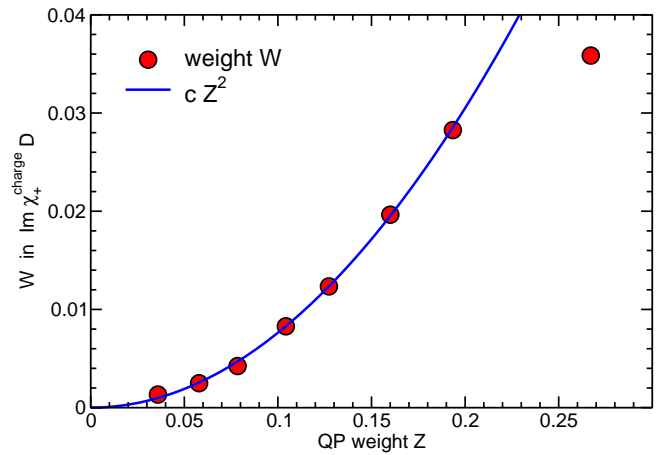}
  \end{center}
  \caption{(color online) Spectral weight $W$ attributed to the
    shoulder in the metallic charge response as function of the
    quasiparticle weight $Z$. The solid line is a quadratic fit with
    $c=0.758$.}
  \label{fig:shoulder-weight}
\end{figure}

In the curves shown in Fig.\ \ref{fig:charge-metal-shoulder} the
shoulder is clearly visible. We extrapolate the main peak on which
the shoulder sits smoothly. This is done by determining a frequency
interval $[b_1,b_2]$ below the shoulder and a second one $[b_3,b_4]$
above it where the shoulder is not present. These intervals are
found from analyzing minima and points of inflections of the
original curve, for example for $U=2.4D$ we took $[0.56D,0.646D]$
and $[1.038D,2.175D]$. Then the data within these two intervals is
interpolated by a 10$^\text{th}$ order polynomial. This is taken to
describe the continuum without the shoulder. The weight of the
shoulder (shaded area in Fig.\ \ref{fig:charge-metal-shoulder}) is
given by integrating the difference between the original curve with
shoulder and the 10$^\text{th}$ order polynomial in the interval
$[b_2,b_3]$. The resulting weights are well-defined within $4\cdot
10^{-4}$.

This procedure is applicable for the results obtained for
$U/D\in\{2.0, 2.2, 2.3 \ldots 2.7, 2.8\}$. The resulting values are
depicted in Fig.\ \ref{fig:shoulder-weight} as function of the
quasiparticle weight $Z$. The quadratic fit agrees very well with
the data except for the last point resulting from $U=2.0D$. For such
a fairly low value of $U$ the separation of the shoulder from its
background is not possible reliably.

Clearly, the shoulder weight $W$ depends quadratically on $Z$. We
recall that the spectral weight $S$ of the sharp feature at the
inner band edges in the single-particle propagator depends linearly
on $Z$: $S\propto Z$ as found previously.\cite{karsk05,karsk08}
These facts are incompatible with the sharp feature $S$ being the
\emph{result} of the shoulder $W$.  It would require that right at
the transition to the insulator the shoulder induces the sharp
feature although the weight of the shoulder is infinitely smaller
than the weight in the sharp peak.

But the other way around the quadratic behavior in Fig.\
\ref{fig:shoulder-weight} finds its natural explanation. The
shoulder results from the convolution of the central quasiparticle
peak of weight $Z$ with the sharp feature with $S\propto Z$. Hence
$W\propto Z^2$ ensues as found.

Also the position in frequency is explained in this way. Since the
central peak is located at zero frequency the shoulder as result of
the convolution with the sharp feature is located at the frequency
where the sharp feature is found.

Summarizing these findings, we conclude that the shoulder in the
charge response can be understood as a \emph{consequence} of the
sharp feature at the inner band edges of the metallic
single-particle spectral density. It is not its cause. While it is
satisfying to have explained the origin of the shoulders in Figs.\
\ref{fig:charge-metal-shoulder} and \ref{fig:shoulder-weight} we
state that the physical origin of the sharp feature described in
Refs.\ \onlinecite{karsk05} and \onlinecite{karsk08} is still
unresolved.


\subsection{Spin Susceptibility}

Here we consider the local observable
\begin{equation}
  \label{eq:qspin}
  Q^\text{spin} = c^\dagger_{i;\uparrow}c^{\phantom\dagger}_{i;\uparrow}
  -c^\dagger_{i;\downarrow}c^{\phantom\dagger}_{i;\downarrow}
\end{equation}
which measures the spin fluctuations around zero magnetization in
$z$ direction.

The general sum rule \eqref{eq:sumrule} requires the expectation
value $\left\langle0\big|Q^2\big|0\right\rangle$ which amounts for
$Q^\text{spin} $ up to an expression which contains again the double
occupancy. First we note
\begin{equation}
  (Q^\text{spin})^2 = \hat n_\uparrow +\hat n_\downarrow -2 
  \hat n_\uparrow \hat n_\downarrow
\end{equation}
where we used that the fermionic occupation number $\hat n$ is equal
to its square. This implies at half-filling
\begin{equation}
  \label{eq:spin-sumrule}
  \int_0^\infty \chi_+^\text{spin}(\omega)d\omega
  = \pi\left(1- 2\left\langle0\big|\hat n_\uparrow \hat n_\downarrow\big|0\right\rangle\right).
\end{equation}
Note that this expression stays finite in the limit of vanishing
double occupancy as it occurs for $U\to\infty$.


\subsubsection{Insulator}

\begin{figure}
  \begin{center}
    \includegraphics[width=0.99\columnwidth]{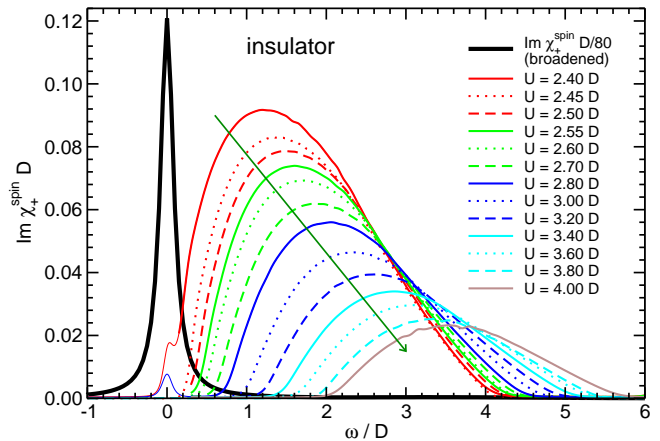}
  \end{center}
  \caption{(color online) Positive imaginary part
    $\chi^\text{spin}_+$ of the local spin susceptibility in the
    insulating phase as function of frequency for various values of
    the interaction $U$. The arrow points in the direction of
    increasing interaction. The Lorentzian at zero frequency
    represents the example at $U=4D$ for the broadened $\delta$ peak
    occurring in the insulator.}
  \label{fig:spin-insulator}
\end{figure}

In Fig.\ \ref{fig:spin-insulator} a series of numerical results is
shown for the positive imaginary part $\chi^\text{spin}_+$ of the
local spin susceptibility in the insulating phase. In the D-DMRG we
kept $m=128$ states in the truncated DMRG basis the mesh is given by
the frequency interval $\Delta\omega=0.05D$, and the imaginary
broadening is $\delta=0.1D$. In the LB deconvolution, the tolerance
constant $A_\text{LB}=1/10$ is used.\cite{raas05a} The curves are
not perfectly smooth but display some very small wiggles due to the
LB deconvolution. The attempt to deconvolve the numerical DMRG data
as a completely continuous spectral density leads to a very large
and very narrow peak at low frequency (not shown). The continua
beside this dominating term cannot be resolved reliably. But it
turns out that the ansatz
\begin{equation}
  \label{eq:A-def}
  \frac{1}{\pi}
  \text{Im}\chi_+^\text{spin} =A\delta(\omega) + \rho^\text{cont}(\omega)
\end{equation}
works extremely well for deconvolution, see Fig.\
\ref{fig:spin-insulator}. Here $\rho^\text{cont}(\omega)$ stands for
the continuous spectral density which is retrieved via the LB
deconvolution. The weight $A$ of the $\delta$ peak results from the
non-linear set of equations defining the Lagrange multipliers
appearing in the LB ansatz.\cite{raas05a}

Why does a zero frequency $\delta$ function make sense physically in
the local spin response of a paramagnetic insulator in infinite
dimensions? The exchange coupling $J$ in a Heisenberg model derived
from a Hubbard model in the insulating regime reads
$J=4t^2/U$.\cite{harri67} Hence scaling $t=t^\star/\sqrt{z}$ implies
$J\propto 1/z$ and the exchange coupling does not contribute unless
all the nearest neighbors contribute on average the same
non-vanishing amount. This implies that a static mean-field
treatment of the Heisenberg antiferromagnet around the Ising limit
becomes exact in infinite dimensions.\cite{klein95} There are no
short-range spin-spin correlations. Hence each spin feels only the
local field $h_\text{MF}=- z J m$ generated by the average
magnetization $m$ of its $z$ neighbors. But in the
\emph{paramagnetic} phase which we consider here the average
magnetization is zero: $m=0$. Hence there is no field and
concomitantly there is no preferred direction of the local
spin. This implies that the spin response is the one of a free spin,
this means, a $\delta$ function at zero frequency. It signals that a
local spin flip does not cost any energy.

\begin{figure}
  \begin{center}
    \includegraphics[width=0.99\columnwidth]{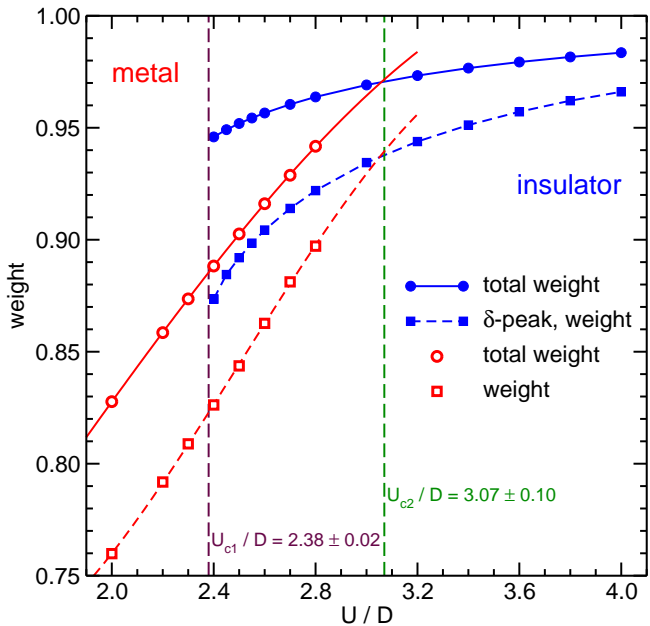}
  \end{center}
  \caption{(color online) The total weight (circles) and four times
    the square of the local $S_z$ component (squares) of the spin
    response $\chi_+^\text{spin}(\omega)$ in the insulating (filled
    symbols) and in the metallic phase (open symbols). In the
    insulator the local $S_z$ component is taken from the weight $A$
    of the $\delta$ peak as in Eq.\ \eqref{eq:A-def}. In the metal
    it is taken to be the weight of the dominant low energy peak,
    see Figs.\ \ref{fig:met-spin-smallU} and
    \ref{fig:met-spin-largeU}. The dark lines interpolate the data
    in the insulator while the light lines interpolate and
    extrapolate the data in the metal.}
  \label{fig:weights}
\end{figure}
Besides understanding this physics on the level of the infinite
dimensional Hubbard model it can be understood on the level of the
effective impurity model. Indeed, a gapped impurity model at
particle-hole symmetry is found to be always in the so-called local
moment regime,\cite{chen98,galpi08,bulla08} where a free local
moment is formed.

The weight $A$ of the $\delta$ function represents four times the
square of the local $S_z$ component in our normalization. It may not
be confused with a static magnetization which is absent in the
paramagnetic phase considered. For $U\to\infty$, $A$ takes the value
unity since the spin $S=1/2$ is fully localized. At any finite
interaction $U<\infty$, $A$ is reduced since charge fluctuations
renormalize its value downward. The spins are not fully localized
but smeared out to some extent to adjacent sites due to their
virtual excursions. This behavior is analyzed quantitatively in
Fig.\ \ref{fig:weights} where the filled squares denote the values
of $A$. Further discussion is presented below.

Besides the $\delta$ peak a continuous contribution of low weight
persists. It correponds to the charge fluctuations which take some
weight away from the dominant local spin response at zero frequency.
Indeed, the continua are very similar, though not identical, to the
ones found in the charge response in the insulating regime, see
Fig.\ \ref{fig:charge-insulator}. To the accuracy that our numerical
deconvolved data allows the onset of the continuous spectral spin
response is again at $2\Delta$, i.e., twice the single-particle gap
in the insulating regime. Hence the origin of the continuous spin
response is essentially the excitation of particle-hole pairs. This
is qualitatively similar to what one expects from the diagrammatic
result in random-phase approximation where the response is a
function of the particle-hole bubble.

The fact that most of the spectral weight in the spin response is
found at zero frequency and not in the continuum starting at
$2\Delta$ shows impressively that a binding phenomenon occurs. The
spin response at zero frequency can be viewed as the signature of a
bound state of a particle-hole pair.\footnote{For this reason one
  could call a collective spin excitation an exciton. But usually
  the latter expression is used only for bound states which are
  weakly bound and do not display binding energies of the order of
  twice the charge gap $2\Delta$.} Only by binding one can
understand how spectral weight can be transferred to energies lower
than the sum of the energies of the constituent states. We think
that such shift of spectral weight due to binding is not taken into
account by the sophisticated argument on the spectral density close
to the Mott transition.\cite{kehre98b} This argument excluded the
continuous scenario at zero temperature which is supported by most
other analytical and numerical evidence \footnote{A notable
  exception are the results in Ref.\ \onlinecite{noack99}.} (see
Sec.\ \ref{sec:intro} for a sketch of this scenario).

The total weight of the spin response as displayed in Fig.\
\ref{fig:weights} just reflects the behavior of the double occupancy
according to the sum rule \eqref{eq:spin-sumrule}. Hence it
approaches unity for $U\to \infty$ but it does not become very small
on $U\to U_{c1}$ either.

Much more interesting is the behavior of the square of the local
$S_z$ component as quantified by $A$ in Eq.\ \eqref{eq:A-def}. The
square root of this expression can be identified with the local
magnetic moment. Clearly, $A=1$ at $U=\infty$ is the starting point.
But it is remarkable that $A\approx 0.94$ has hardly decreased for
$U=U_{c2}$ where the insulator is no longer the ground state. In
physical terms this means that a Mott insulator is governed by very
well localized spins as long as it exists. Hardly any
renormalization due to charge fluctuations takes place. From a
theoretical point of view this can be explained by the significant
charge gap $\Delta\approx 0.45 D$ at $U_{c2}$ (Ref.\
\onlinecite{karsk05}) which acts as an infrared cutoff limiting the
influence of charge fluctuations. This can be easily understood by
the renormalization flow of the impurity model where the insulator
corresponds to the local moment fixed point.\cite{chen98,bulla08}

Even more remarkable is that the local magnetic moment is not much
lower at $U_{c1}$ either. Below this interaction the insulator is
not longer locally stable. Even there $A$ is still larger than about
$0.87$ although the charge gap has become zero and the lower and the
upper Hubbard band are touching each other, see for instance Fig.\ 2
in Ref.\ \onlinecite{karsk05}. But it is obvious from the touching
Hubbard bands that the DOS $\rho(\omega)$ at the Fermi level, i.e.,
at $\omega=0$, is still zero. We conclude that no hard infrared
cutoff is needed and that the fact that the insulator at $U_{c1}$
displays a semi-metallic DOS with $\lim_{\omega\to 0}\rho(\omega)=0$
is sufficient to bound the influence of the charge fluctuations. So
the magnetic moment is not renormalized to zero and the fixed point
of the renormalization of the corresponding impurity model is still
the local moment fixed point.\cite{chen98,bulla08}

The main goal of this paper is the comprehensive analy\-sis of the
susceptibility around a Mott transition in infinite dimensions. But
in view of the remarkable findings at $d=\infty$ it is in order to
speculate how these findings change on passing to finite dimensional
systems.

The main difference is that any finite dimensional system would show
at least short-range magnetic correlations. Hence the magnetic
response would not be governed by a $\delta$ peak at zero frequency
as in Fig.\ \ref{fig:spin-insulator}. If the resulting
antiferromagnetic system is sufficiently strongly frustrated and/or
sufficiently low-dimensional so that the magnetic fluctuations are
strong enough to prevent magnetic long-range order the system would
be paramagnetic displaying a magnetic gap. Generically, the magnetic
excitations would be triplons\cite{schmi03c} with some
dispersion. Hence the local spin response would show the sum of
triplon contributions from all the wave vectors in the Brillouin
zone. A broad feature at finite frequencies in a frequency range
given by the magnetic exchange $J$ would be seen in
$\text{Im}\chi_+^\text{spin}(\omega)$. A sharp mode at finite, but
low frequency would be discernible in
$\text{Im}\chi_+^\text{spin}(\omega,\mathbf{q})$ at a given wave
vector $\mathbf{q}$. The sum of the weights in these sharp modes
over the Brillouin zone constitutes the finite dimensional analog of
the weight $A$ in our infinite dimensional analy\-sis. We expect
other features to be qualitatively very similar to the above
findings. For instance the local magnetic moment in any insulating
state should be very little renormalized due to charge fluctuations.


\subsubsection{Metal}

\begin{figure}
  \begin{center}
    \includegraphics[width=0.99\columnwidth]{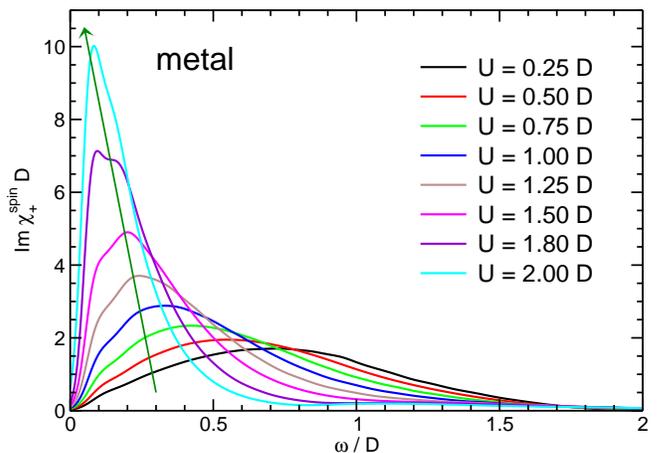}
  \end{center}
  \caption{(color online) Positive imaginary part
    $\chi^\text{spin}_+$ of the local spin susceptibility in the
    metallic phase as function of frequency for various small values
    of the interaction $U$. The arrow points in the direction of
    increasing interaction. Note the quickly increasing peak at low
    frequencies and the quickly vanishing spectral weight at higher
    frequencies.}
  \label{fig:met-spin-smallU}
\end{figure}
\begin{figure}
  \begin{center}
    \includegraphics[width=0.99\columnwidth]{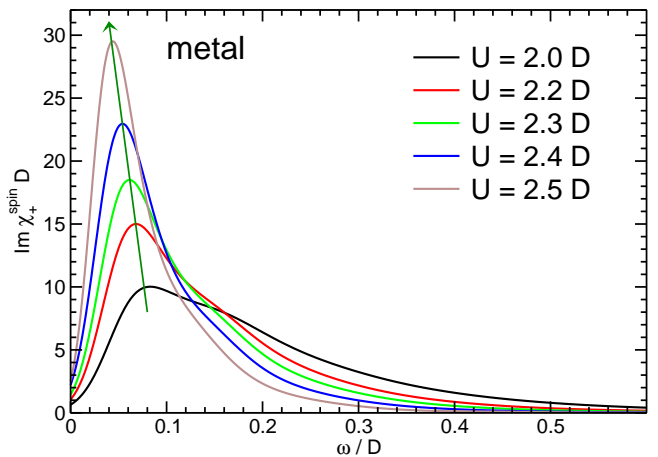}
  \end{center}
  \caption{(color online) Positive imaginary part
    $\chi^\text{spin}_+$ of the local spin susceptibility in the
    metallic phase as function of frequency for various values of
    the interaction $U$ close to the transition to the insulator.
    The arrow points in the direction of increasing interaction. The
    peak at low frequencies prevails completely, note the scale on
    the $y$ axis.}
  \label{fig:met-spin-largeU}
\end{figure}

In Fig.\ \ref{fig:met-spin-smallU} a series of numerical results is
shown for the positive imaginary part $\chi^\text{spin}_+$ of the
local spin susceptibility in the metallic phase for not too large
values of the interaction. The curves for larger values of $U$ are
plotted in Fig.\ \ref{fig:met-spin-largeU} In the D-DMRG we kept
between $m=128$ and $m=256$ states in the truncated DMRG basis, the
frequency mesh is given by the intervals $\Delta\omega=0.025D$ and
$\Delta\omega=0.05D$, and the imaginary broadening is chosen between
$\delta=0.05D$ and $\delta=0.1D$. The LB deconvolution\cite{raas05a}
is done with the tolerance constant $1/A_\text{LB}=1/10$.

\begin{figure}
  \begin{center}
   \includegraphics[width=0.99\columnwidth]{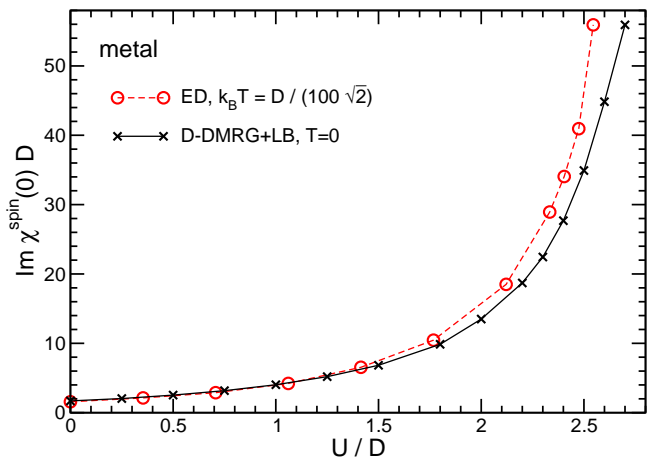}
  \end{center}
  \caption{(color online) Static susceptibility
    $\chi^\text{spin}(0)$ from our dynamic data via \eqref{eq:KK}
    (black crosses) compared to data obtained by exact
    diagonalization at very small temperature (red circles), adapted
    from Fig.\ 44 in Ref.\ \onlinecite{georg96}. The black cross at
    $U=0$ corresponds to the analytic result $\chi^\text{spin}(0)
    =16/(3\pi D)$. Lines are guides to the eye only.}
  \label{fig:gauge}
\end{figure}

As a first check of our data we compute the static spin spin
susceptibility $\chi^\text{spin}(0)$ via the Kramers-Kronig relation
\begin{equation}
  \label{eq:KK}
  \chi^\text{spin}(0) = 
  \frac{2}{\pi}\int_0^\infty 
  \frac{\text{Im}\chi_+^\text{spin}(\omega)-
    \text{Im}\chi_+^\text{spin}(-\omega)}{\omega}d\omega .
\end{equation}
The negative term in the numerator occurs only because the LB
deconvolution tends to produce spurious minor contributions at
negative frequencies; otherwise
$\text{Im}\chi_+^\text{spin}(\omega<0)$ is strictly zero at zero
temperature. The results are compared to previous results in Fig.\
\ref{fig:gauge}, see Fig.\ 44 in Ref.\ \onlinecite{georg96}, which
were obtained by exact diagonalization. The agreement is very good
at low $U$ deteriorating for larger values of $U$. We attribute the
discrepancy at larger values of $U$ to effects of finite size and of
finite temperature in the exact diagonalization approach. But for
small and intermediate values of the interaction the comparison
underlines the validity of our results.

The curves are dominated by a prominent peak at low frequencies. In
some curves, in particular between $U=1D$ and $2D$, there appears to
be a shoulder to this peak. Since this feature occurs already for
moderate values of $U$ where the curves are still fairly smooth we
are confident that this shoulder is in fact a real physical
feature. But due to the deconvolution procedure involved we cannot
at present rule out completely that it is a spurious numerical
effect. In the following we refrain from discussing the shape of the
low-frequency peak.

\begin{figure}
  \begin{center}
    \includegraphics[width=0.99\columnwidth]{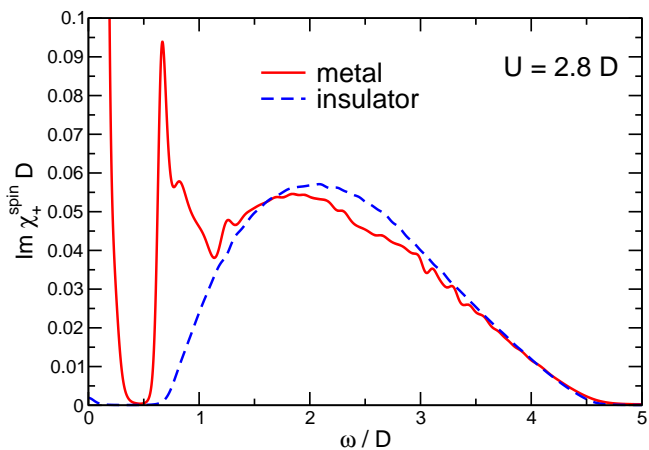}
  \end{center}
  \caption{(color online) Positive imaginary part
    $\chi^\text{spin}_+$ of the local spin susceptibility as
    function of frequency for interaction $U=2.8D$ in the metallic
    and insulating phase.}
  \label{fig:spin-compare}
\end{figure}

At higher frequencies around $\omega\approx 2D$ in Fig.\
\ref{fig:met-spin-smallU} a very broad peak of low amplitude can be
seen. Qualitatively, this continuum results from the convolution of
the lower and the upper Hubbard band or more complicated
descriptions in terms of excitations from the lower and the upper
Hubbard band.\cite{logan98,galpi08} It corresponds to the continuum
found in the insulating phase in Fig.\ \ref{fig:spin-insulator}.

The insulating and the metallic response at higher frequencies are
alike, see Fig.\ \ref{fig:spin-compare}, because their
single-particle spectral densities are identical for higher
frequencies for $U\to U_{c2}$, see Fig.\ 2 in Ref.\
\onlinecite{karsk05}. This is analogous to what we have discussed in
the charge response in Fig.\ \ref{fig:charge-compare}. In both
cases, the insulating and the metallic responses coincide for
$\omega\gtrapprox 3D$. The slightly more wiggly metallic spin
response in Fig.\ \ref{fig:spin-compare} results from the difficulty
to deconvolve the relatively small continuum close to the dominating
low-frequency peak.

Since the low-frequency peak is clearly the dominating feature its
physical significance has to be elucidated. Clearly, it is shifted
towards zero frequency on $U\to U_{c2}$ while becoming narrower and
narrower. So it is to be expected that it represents the precursor
in the metal of the $\delta$ peak in the insulating regime. In order
to support this claim we analyze the sum rule
\eqref{eq:spin-sumrule} and the weight in the low-frequency
peak. Both sets of data are depicted for the metal in Fig.\
\ref{fig:weights} by the open symbols. The sum rule is again well
fulfilled to within a absolute relative error of $0.5\%$. The
weight in the low-frequency peak is integrated till the first
minimum in the spectral density on the right hand side of the peak
is reached. First, we note that the total spectral weight of the
metal equals the one of the insulator for $U\to U_{c2}$ as far as
the extrapolation can be trusted. This is expected since the
transition occurs precisely where the double occupancies become
equal.\cite{karsk05,blume05a}

Second, we note that also the weight of the dominant low-frequency
peak in the metal approaches the weight of the $\delta$ peak in the
insulator to very good accuracy. This clearly corroborates our
hypothesis that the metallic dominant low-frequency peak is the
precursor of the bound state at zero frequency in the
insulator. Naturally, there is no sharp bound state in the metal
because such a state can decay into particle-hole states made from
the heavy quasiparticles and -holes which are still present in the
metal. Hence no bound state but a resonance occurs. The width of
this resonance can be understood as Landau damping.

Recall that the low-energy spin resonance in the gapless case
becomes the zero-frequency mode of the corresponding gapped case in
single impurity Anderson models.\cite{galpi08} So our analysis is
well-founded also on the level of the impurity models.

In the insulating regime we have tentatively carried our infinite
dimensional results over to finite dimensions. We have speculated
that the $\delta$ peak in Fig.\ \ref{fig:spin-insulator} becomes a
dispersive magnon, if magnetic long-range order exists, or a
dispersive triplon if not. Both, magnon or triplon, are bound states
of particle-hole pairs from the electronic point of view. So we
expect that the emergent magnetic resonance found here becomes in
finite dimensions a dispersive resonance which is the precursor of a
perfectly sharp magnetic excitation. In literature, the term
`paramagnon' is used for such precursive resonances. If no magnetic
order is to be expected the term `paratriplon' would be more
appropriate.

We recall that the existence of precursive magnetic excitations and
their interaction with the single-particle excitations is very
important for the understanding of kinks in the electronic
dispersions\cite{raas09a,byczu07a} and possibly also for Cooper
pairing in strongly interacting systems.


\section{Conclusions}

In this paper, we investigated the zero temperature Mott transition
as function of the interaction in a generic model, namely the
half-filled Hubbard model in infinite dimensions. We focused on the
susceptibilities which are of theoretical and experimental
relevance. Thereby, complementary information to the existing
investigations of the single-particle dynamics is provided.

We showed that in infinite dimensions the generic susceptibilities
are the local ones. Locally, only three types of bosonic observables
exist: the charge, the spin and the Cooper pairing operator. So we
discussed the corresponding susceptibilites $\chi^\text{charge}$,
$\chi^\text{spin}$, and $\chi^\text{pair}$. By an intricate
symmetry, $\chi^\text{pair}$ is found to be identical to
$\chi^\text{charge}$.

For the charge susceptibility $\chi^\text{charge}$ in the insulating
phase we found a strong suppresion on increasing repulsive
interaction. The spectral density sets in at $2\Delta$, i.e., at
twice the single-particle gap. No binding phenomenon occurs; the
spectral line is rather featureless.

In the metallic phase, three ranges in frequency can be
distinguished. The first results from a heavy quasiparticle and a
heavy quasihole, the second from one heavy excitation and one in one
of the two Hubbard bands, and the third consists of a particle in
the upper and a hole in the lower Hubbard band. On approaching the
insulator $U\to U_{c2}$ the quasiparticle weight $Z$ vanishes
linearly.\cite{moell95,bulla99a,karsk05} The weight in the first
region vansihes like $Z^2$, the weight in the second region like $Z$
while the weight in the third region, though small, persists. It is
also found in the insulator.

A shoulder occurs in the metallic charge response at the same
energies as the sharp feature found previously at the inner band
edges.\cite{karsk05,karsk08} The weight in the shoulder scales like
$Z^2$ so that we are led to the conclusion that the shoulder is a
consequence rather than the cause of the sharp feature in the
single-particle propagator.

The spin susceptibility $\chi^\text{spin}$ in the insulating phase
is found to be dominated by a strong $\delta$ peak which would
correspond in finite dimensions to dispersive magnetic excitations.
In infinite dimensions in a paramagnetic insulator it happens to be
at zero frequency. The peak must be seen as a particle-hole bound
state.

Besides this peak only a very weak continuum is found at higher
frequencies. Hence the localized magnetic moment is only very weakly
reduced by charge excitations. The Mott insulator is governed by
very well localized spins as long as it exists.

In the metallic phase, the spin response at higher frequencies
displays again only a continuum of small spectral weight. Upon
increasing interaction the spin spectral density is dominated by a
pronounced peak at low, but finite, frequencies. This peak comprises
most of the spectral weight. It constitutes the precursor of the
sharp magnetic mode in the insulator as is evidenced by the
coinciding spectral weights of both features at $U=U_{c2}$. The
pronounced metallic peak is the signature of an almost bound
particle-hole resonance which can be seen as the emergent magnetic
mode (paramagnon or paratriplon). We expect this mode to persist in
finite dimensions as a dispersive resonance at low, but finite
frequencies.

This concludes the investigation of the zero temperature Mott
transition at half-filling in infinite dimensions. Further
investigations away from half-filling are called for. Similarly, it
would be very important to verify the hypotheses derived here for
finite dimensions by future calculations.

\acknowledgments

We like to thank M.\ Karski for providing data and F.B.\ Anders and
D. Vollhardt for very helpful discussions. Financial support by the
Heinrich-Hertz Stiftung des Landes Nordrhein-Westfalen is gratefully
acknowledged by one of us (GSU).


\begin{thebibliography}{47}
\expandafter\ifx\csname natexlab\endcsname\relax\def\natexlab#1{#1}\fi
\expandafter\ifx\csname bibnamefont\endcsname\relax
  \def\bibnamefont#1{#1}\fi
\expandafter\ifx\csname bibfnamefont\endcsname\relax
  \def\bibfnamefont#1{#1}\fi
\expandafter\ifx\csname citenamefont\endcsname\relax
  \def\citenamefont#1{#1}\fi
\expandafter\ifx\csname url\endcsname\relax
  \def\url#1{\texttt{#1}}\fi
\expandafter\ifx\csname urlprefix\endcsname\relax\def\urlprefix{URL }\fi
\providecommand{\bibinfo}[2]{#2}
\providecommand{\eprint}[2][]{\url{#2}}

\bibitem[{\citenamefont{Hubbard}(1963)}]{hubba63}
\bibinfo{author}{\bibfnamefont{J.}~\bibnamefont{Hubbard}},
  \bibinfo{journal}{Proc. R. Soc. London, Ser. A} \textbf{\bibinfo{volume}{276}},
  \bibinfo{pages}{238} (\bibinfo{year}{1963}).

\bibitem[{\citenamefont{Kanamori}(1963)}]{kanam63}
\bibinfo{author}{\bibfnamefont{J.}~\bibnamefont{Kanamori}},
  \bibinfo{journal}{Prog. Theor. Phys.} \textbf{\bibinfo{volume}{30}},
  \bibinfo{pages}{275} (\bibinfo{year}{1963}).

\bibitem[{\citenamefont{Gutzwiller}(1963)}]{gutzw63}
\bibinfo{author}{\bibfnamefont{M.~C.} \bibnamefont{Gutzwiller}},
  \bibinfo{journal}{Phys. Rev. Lett.} \textbf{\bibinfo{volume}{10}},
  \bibinfo{pages}{159} (\bibinfo{year}{1963}).

\bibitem[{\citenamefont{Essler et~al.}(2005)\citenamefont{Essler, Frahm,
  G\"ohmann, Kl\"umper, and Korepin}}]{essle05}
\bibinfo{author}{\bibfnamefont{F.~H.~L.} \bibnamefont{Essler}},
  \bibinfo{author}{\bibfnamefont{H.}~\bibnamefont{Frahm}},
  \bibinfo{author}{\bibfnamefont{F.}~\bibnamefont{G\"ohmann}},
  \bibinfo{author}{\bibfnamefont{A.}~\bibnamefont{Kl\"umper}},
  \bibnamefont{and} \bibinfo{author}{\bibfnamefont{V.~E.}
  \bibnamefont{Korepin}}, \emph{\bibinfo{title}{The One-Dimensional Hubbard
  Model}} (\bibinfo{publisher}{Cambridge University Press},
  \bibinfo{address}{Cambridge, United Kingdom}, \bibinfo{year}{2005}).

\bibitem[{\citenamefont{Metzner and Vollhardt}(1989)}]{metzn89a}
\bibinfo{author}{\bibfnamefont{W.}~\bibnamefont{Metzner}} \bibnamefont{and}
  \bibinfo{author}{\bibfnamefont{D.}~\bibnamefont{Vollhardt}},
  \bibinfo{journal}{Phys. Rev. Lett.} \textbf{\bibinfo{volume}{62}},
  \bibinfo{pages}{324} (\bibinfo{year}{1989}).

\bibitem[{\citenamefont{M\"uller-Hartmann}(1989)}]{mulle89a}
\bibinfo{author}{\bibfnamefont{E.}~\bibnamefont{M\"uller-Hartmann}},
  \bibinfo{journal}{Z. Phys. B} \textbf{\bibinfo{volume}{74}},
  \bibinfo{pages}{507} (\bibinfo{year}{1989}).

\bibitem[{\citenamefont{Pruschke et~al.}(1995)\citenamefont{Pruschke, Jarrell,
  and Freericks}}]{prusc95}
\bibinfo{author}{\bibfnamefont{T.}~\bibnamefont{Pruschke}},
  \bibinfo{author}{\bibfnamefont{M.}~\bibnamefont{Jarrell}}, \bibnamefont{and}
  \bibinfo{author}{\bibfnamefont{J.~K.} \bibnamefont{Freericks}},
  \bibinfo{journal}{Adv. Phys.} \textbf{\bibinfo{volume}{44}},
  \bibinfo{pages}{187} (\bibinfo{year}{1995}).

\bibitem[{\citenamefont{Georges et~al.}(1996)\citenamefont{Georges, Kotliar,
  Krauth, and Rozenberg}}]{georg96}
\bibinfo{author}{\bibfnamefont{A.}~\bibnamefont{Georges}},
  \bibinfo{author}{\bibfnamefont{G.}~\bibnamefont{Kotliar}},
  \bibinfo{author}{\bibfnamefont{W.}~\bibnamefont{Krauth}}, \bibnamefont{and}
  \bibinfo{author}{\bibfnamefont{M.~J.} \bibnamefont{Rozenberg}},
  \bibinfo{journal}{Rev. Mod. Phys.} \textbf{\bibinfo{volume}{68}},
  \bibinfo{pages}{13} (\bibinfo{year}{1996}).

\bibitem[{\citenamefont{Bulla}(1999)}]{bulla99a}
\bibinfo{author}{\bibfnamefont{R.}~\bibnamefont{Bulla}},
  \bibinfo{journal}{Phys. Rev. Lett.} \textbf{\bibinfo{volume}{83}},
  \bibinfo{pages}{136} (\bibinfo{year}{1999}).

\bibitem[{\citenamefont{Bulla et~al.}(2001)\citenamefont{Bulla, Costi, and
  Vollhardt}}]{bulla01a}
\bibinfo{author}{\bibfnamefont{R.}~\bibnamefont{Bulla}},
  \bibinfo{author}{\bibfnamefont{T.~A.} \bibnamefont{Costi}}, \bibnamefont{and}
  \bibinfo{author}{\bibfnamefont{D.}~\bibnamefont{Vollhardt}},
  \bibinfo{journal}{Phys. Rev. B} \textbf{\bibinfo{volume}{64}},
  \bibinfo{pages}{045103} (\bibinfo{year}{2001}).

\bibitem[{\citenamefont{Bl\"umer}(2002)}]{blume02}
\bibinfo{author}{\bibfnamefont{N.}~\bibnamefont{Bl\"umer}},
  \emph{\bibinfo{title}{Mott-Hubbard Metal-Insulator Transition and Optical
  Conductivity in High Dimensions}} (\bibinfo{publisher}{PhD thesis},
  \bibinfo{address}{Universit\"at Augsburg, Germany}, \bibinfo{year}{2002}).

\bibitem[{\citenamefont{Potthoff}(2003)}]{potth03b}
\bibinfo{author}{\bibfnamefont{M.}~\bibnamefont{Potthoff}},
  \bibinfo{journal}{Eur. Phys. J. B} \textbf{\bibinfo{volume}{36}},
  \bibinfo{pages}{335} (\bibinfo{year}{2003}).

\bibitem[{\citenamefont{Karski et~al.}(2005)\citenamefont{Karski, Raas, and
  Uhrig}}]{karsk05}
\bibinfo{author}{\bibfnamefont{M.}~\bibnamefont{Karski}},
  \bibinfo{author}{\bibfnamefont{C.}~\bibnamefont{Raas}}, \bibnamefont{and}
  \bibinfo{author}{\bibfnamefont{G.~S.} \bibnamefont{Uhrig}},
  \bibinfo{journal}{Phys. Rev. B} \textbf{\bibinfo{volume}{72}},
  \bibinfo{pages}{113110} (\bibinfo{year}{2005}).

\bibitem[{\citenamefont{Karski et~al.}(2008)\citenamefont{Karski, Raas, and
  Uhrig}}]{karsk08}
\bibinfo{author}{\bibfnamefont{M.}~\bibnamefont{Karski}},
  \bibinfo{author}{\bibfnamefont{C.}~\bibnamefont{Raas}}, \bibnamefont{and}
  \bibinfo{author}{\bibfnamefont{G.~S.} \bibnamefont{Uhrig}},
  \bibinfo{journal}{Phys. Rev. B} \textbf{\bibinfo{volume}{77}},
  \bibinfo{pages}{075116} (\bibinfo{year}{2008}).

\bibitem[{\citenamefont{Brinkman and Rice}(1970)}]{brink70}
\bibinfo{author}{\bibfnamefont{W.~F.} \bibnamefont{Brinkman}} \bibnamefont{and}
  \bibinfo{author}{\bibfnamefont{T.~M.} \bibnamefont{Rice}},
  \bibinfo{journal}{Phys. Rev. B} \textbf{\bibinfo{volume}{2}},
  \bibinfo{pages}{1324} (\bibinfo{year}{1970}).

\bibitem[{\citenamefont{Hubbard}(1964)}]{hubba64b}
\bibinfo{author}{\bibfnamefont{J.}~\bibnamefont{Hubbard}},
  \bibinfo{journal}{Proc. R. Soc. London, Ser. A} \textbf{\bibinfo{volume}{281}},
  \bibinfo{pages}{401} (\bibinfo{year}{1964}).

\bibitem[{\citenamefont{Zhang et~al.}(1993)\citenamefont{Zhang, Rozenberg, and
  Kotliar}}]{zhang93}
\bibinfo{author}{\bibfnamefont{X.~Y.} \bibnamefont{Zhang}},
  \bibinfo{author}{\bibfnamefont{M.~J.} \bibnamefont{Rozenberg}},
  \bibnamefont{and} \bibinfo{author}{\bibfnamefont{G.}~\bibnamefont{Kotliar}},
  \bibinfo{journal}{Phys. Rev. Lett.} \textbf{\bibinfo{volume}{70}},
  \bibinfo{pages}{1666} (\bibinfo{year}{1993}).

\bibitem[{\citenamefont{Kotliar}(1999)}]{kotli99}
\bibinfo{author}{\bibfnamefont{G.}~\bibnamefont{Kotliar}},
  \bibinfo{journal}{Eur. Phys. J. B} \textbf{\bibinfo{volume}{11}},
  \bibinfo{pages}{27} (\bibinfo{year}{1999}).

\bibitem[{\citenamefont{Byczuk et~al.}(2007)\citenamefont{Byczuk, Kollar, Held,
  Yang, Nekrasov, Pruschke, and Vollhardt}}]{byczu07a}
\bibinfo{author}{\bibfnamefont{K.}~\bibnamefont{Byczuk}},
  \bibinfo{author}{\bibfnamefont{M.}~\bibnamefont{Kollar}},
  \bibinfo{author}{\bibfnamefont{K.}~\bibnamefont{Held}},
  \bibinfo{author}{\bibfnamefont{Y.-F.} \bibnamefont{Yang}},
  \bibinfo{author}{\bibfnamefont{I.~A.} \bibnamefont{Nekrasov}},
  \bibinfo{author}{\bibfnamefont{T.}~\bibnamefont{Pruschke}}, \bibnamefont{and}
  \bibinfo{author}{\bibfnamefont{D.}~\bibnamefont{Vollhardt}},
  \bibinfo{journal}{Nature Phys.} \textbf{\bibinfo{volume}{3}},
  \bibinfo{pages}{168} (\bibinfo{year}{2007}).

\bibitem[{\citenamefont{Raas et~al.}(2008)\citenamefont{Raas, Grete, and
  Uhrig}}]{raas09a}
\bibinfo{author}{\bibfnamefont{C.}~\bibnamefont{Raas}},
  \bibinfo{author}{\bibfnamefont{P.}~\bibnamefont{Grete}}, \bibnamefont{and}
  \bibinfo{author}{\bibfnamefont{G.~S.} \bibnamefont{Uhrig}},
  \bibinfo{journal}{Phys. Rev. Lett.} \textbf{\bibinfo{volume}{102}},
  \bibinfo{pages}{076406} (\bibinfo{year}{2009}).

\bibitem[{\citenamefont{Nishimoto et~al.}(2006)\citenamefont{Nishimoto,
  Gebhard, and Jeckelmann}}]{nishi06}
\bibinfo{author}{\bibfnamefont{S.}~\bibnamefont{Nishimoto}},
  \bibinfo{author}{\bibfnamefont{F.}~\bibnamefont{Gebhard}}, \bibnamefont{and}
  \bibinfo{author}{\bibfnamefont{E.}~\bibnamefont{Jeckelmann}},
  \bibinfo{journal}{Physica B} \textbf{\bibinfo{volume}{378-380}},
  \bibinfo{pages}{283} (\bibinfo{year}{2006}).

\bibitem[{\citenamefont{\v{Z}itko and Pruschke}(2009)}]{zitko09}
\bibinfo{author}{\bibfnamefont{R.}~\bibnamefont{\v{Z}itko}} \bibnamefont{and}
  \bibinfo{author}{\bibfnamefont{T.}~\bibnamefont{Pruschke}},
  \bibinfo{journal}{Phys. Rev. B} \textbf{\bibinfo{volume}{79}},
  \bibinfo{pages}{085106} (\bibinfo{year}{2009}).

\bibitem[{\citenamefont{Bl\"umer}(2008)}]{blume08}
\bibinfo{author}{\bibfnamefont{N.}~\bibnamefont{Bl\"umer}},
  \bibinfo{journal}{arXiv:0801.1222}  (\bibinfo{year}{2008}).

\bibitem[{\citenamefont{Georges and Kotliar}(1992)}]{georg92a}
\bibinfo{author}{\bibfnamefont{A.}~\bibnamefont{Georges}} \bibnamefont{and}
  \bibinfo{author}{\bibfnamefont{G.}~\bibnamefont{Kotliar}},
  \bibinfo{journal}{Phys. Rev. B} \textbf{\bibinfo{volume}{45}},
  \bibinfo{pages}{6479} (\bibinfo{year}{1992}).

\bibitem[{\citenamefont{Jarrell}(1992)}]{jarre92}
\bibinfo{author}{\bibfnamefont{M.}~\bibnamefont{Jarrell}},
  \bibinfo{journal}{Phys. Rev. Lett.} \textbf{\bibinfo{volume}{69}},
  \bibinfo{pages}{168} (\bibinfo{year}{1992}).

\bibitem[{\citenamefont{Economou}(1979)}]{econo79}
\bibinfo{author}{\bibfnamefont{E.~N.} \bibnamefont{Economou}},
  \emph{\bibinfo{title}{Green's Functions in Quantum Physics}},
  vol.~\bibinfo{volume}{7} of \emph{\bibinfo{series}{Solid State Sciences}}
  (\bibinfo{publisher}{Springer}, \bibinfo{address}{Berlin},
  \bibinfo{year}{1979}).

\bibitem[{\citenamefont{Hewson}(1993)}]{hewso93}
\bibinfo{author}{\bibfnamefont{A.~C.} \bibnamefont{Hewson}},
  \emph{\bibinfo{title}{The Kondo Problem to Heavy Fermions}}
  (\bibinfo{publisher}{Cambridge University Press},
  \bibinfo{address}{Cambridge}, \bibinfo{year}{1993}).

\bibitem[{\citenamefont{Garcia et~al.}(2004)\citenamefont{Garcia, Hallberg, and
  Rozenberg}}]{garci04}
\bibinfo{author}{\bibfnamefont{D.~J.} \bibnamefont{Garcia}},
  \bibinfo{author}{\bibfnamefont{K.}~\bibnamefont{Hallberg}}, \bibnamefont{and}
  \bibinfo{author}{\bibfnamefont{M.~J.} \bibnamefont{Rozenberg}},
  \bibinfo{journal}{Phys. Rev. Lett.} \textbf{\bibinfo{volume}{93}},
  \bibinfo{pages}{246403} (\bibinfo{year}{2004}).

\bibitem[{\citenamefont{Raas et~al.}(2004)\citenamefont{Raas, Uhrig, and
  Anders}}]{raas04a}
\bibinfo{author}{\bibfnamefont{C.}~\bibnamefont{Raas}},
  \bibinfo{author}{\bibfnamefont{G.~S.} \bibnamefont{Uhrig}}, \bibnamefont{and}
  \bibinfo{author}{\bibfnamefont{F.~B.} \bibnamefont{Anders}},
  \bibinfo{journal}{Phys. Rev. B} \textbf{\bibinfo{volume}{69}},
  \bibinfo{pages}{041102(R)} (\bibinfo{year}{2004}).

\bibitem[{\citenamefont{Uhrig and Vlaming}(1993)}]{uhrig93b}
\bibinfo{author}{\bibfnamefont{G.~S.} \bibnamefont{Uhrig}} \bibnamefont{and}
  \bibinfo{author}{\bibfnamefont{R.}~\bibnamefont{Vlaming}},
  \bibinfo{journal}{Phys. Rev. Lett.} \textbf{\bibinfo{volume}{71}},
  \bibinfo{pages}{271} (\bibinfo{year}{1993}).

\bibitem[{\citenamefont{Ramasesha et~al.}(1997)\citenamefont{Ramasesha, Pati,
  Krishnamurthy, Shuai, and Br\'{e}das}}]{ramas97}
\bibinfo{author}{\bibfnamefont{S.}~\bibnamefont{Ramasesha}},
  \bibinfo{author}{\bibfnamefont{S.~K.} \bibnamefont{Pati}},
  \bibinfo{author}{\bibfnamefont{H.~R.} \bibnamefont{Krishnamurthy}},
  \bibinfo{author}{\bibfnamefont{Z.}~\bibnamefont{Shuai}}, \bibnamefont{and}
  \bibinfo{author}{\bibfnamefont{J.~L.} \bibnamefont{Br\'{e}das}},
  \bibinfo{journal}{Synthetic Metals} \textbf{\bibinfo{volume}{85}},
  \bibinfo{pages}{1019} (\bibinfo{year}{1997}).

\bibitem[{\citenamefont{K\"uhner and Monien}(1998)}]{kuhne98}
\bibinfo{author}{\bibfnamefont{T.~D.} \bibnamefont{K\"uhner}} \bibnamefont{and}
  \bibinfo{author}{\bibfnamefont{H.}~\bibnamefont{Monien}},
  \bibinfo{journal}{Phys. Rev. B} \textbf{\bibinfo{volume}{58}},
  \bibinfo{pages}{R14741} (\bibinfo{year}{1998}).

\bibitem[{\citenamefont{Jeckelmann}(2002)}]{jecke02}
\bibinfo{author}{\bibfnamefont{E.}~\bibnamefont{Jeckelmann}},
  \bibinfo{journal}{Phys. Rev. B} \textbf{\bibinfo{volume}{66}},
  \bibinfo{pages}{045114} (\bibinfo{year}{2002}).

\bibitem[{\citenamefont{Raas and Uhrig}(2005)}]{raas05a}
\bibinfo{author}{\bibfnamefont{C.}~\bibnamefont{Raas}} \bibnamefont{and}
  \bibinfo{author}{\bibfnamefont{G.~S.} \bibnamefont{Uhrig}},
  \bibinfo{journal}{Eur. Phys. J. B} \textbf{\bibinfo{volume}{45}},
  \bibinfo{pages}{293} (\bibinfo{year}{2005}).

\bibitem[{\citenamefont{Demler et~al.}(2004)\citenamefont{Demler, Hanke, and
  Zhang}}]{demle04}
\bibinfo{author}{\bibfnamefont{E.}~\bibnamefont{Demler}},
  \bibinfo{author}{\bibfnamefont{W.}~\bibnamefont{Hanke}}, \bibnamefont{and}
  \bibinfo{author}{\bibfnamefont{S.-C.} \bibnamefont{Zhang}},
  \bibinfo{journal}{Rev. Mod. Phys.} \textbf{\bibinfo{volume}{76}},
  \bibinfo{pages}{909} (\bibinfo{year}{2004}).

\bibitem[{\citenamefont{Nishimoto et~al.}(2004)\citenamefont{Nishimoto,
  Gebhard, and Jeckelmann}}]{nishi04b}
\bibinfo{author}{\bibfnamefont{S.}~\bibnamefont{Nishimoto}},
  \bibinfo{author}{\bibfnamefont{F.}~\bibnamefont{Gebhard}}, \bibnamefont{and}
  \bibinfo{author}{\bibfnamefont{E.}~\bibnamefont{Jeckelmann}},
  \bibinfo{journal}{J. Phys.: Condens. Matter} \textbf{\bibinfo{volume}{16}},
  \bibinfo{pages}{7063} (\bibinfo{year}{2004}).

\bibitem[{\citenamefont{Logan et~al.}(1998)\citenamefont{Logan, Eastwood, and
  Tusch}}]{logan98}
\bibinfo{author}{\bibfnamefont{D.~E.} \bibnamefont{Logan}},
  \bibinfo{author}{\bibfnamefont{M.~P.} \bibnamefont{Eastwood}},
  \bibnamefont{and} \bibinfo{author}{\bibfnamefont{M.~A.} \bibnamefont{Tusch}},
  \bibinfo{journal}{J. Phys.: Condens. Matter} \textbf{\bibinfo{volume}{10}},
  \bibinfo{pages}{2673} (\bibinfo{year}{1998}).

\bibitem[{\citenamefont{Galpin and Logan}(2008)}]{galpi08}
\bibinfo{author}{\bibfnamefont{M.~R.} \bibnamefont{Galpin}} \bibnamefont{and}
  \bibinfo{author}{\bibfnamefont{D.~E.} \bibnamefont{Logan}},
  \bibinfo{journal}{Eur. Phys. J. B} \textbf{\bibinfo{volume}{62}},
  \bibinfo{pages}{129} (\bibinfo{year}{2008}).

\bibitem[{\citenamefont{Harris and Lange}(1967)}]{harri67}
\bibinfo{author}{\bibfnamefont{A.~B.} \bibnamefont{Harris}} \bibnamefont{and}
  \bibinfo{author}{\bibfnamefont{R.~V.} \bibnamefont{Lange}},
  \bibinfo{journal}{Phys. Rev.} \textbf{\bibinfo{volume}{157}},
  \bibinfo{pages}{295} (\bibinfo{year}{1967}).

\bibitem[{\citenamefont{Kleine et~al.}(1995)\citenamefont{Kleine, Uhrig, and
  M\"uller-Hartmann}}]{klein95}
\bibinfo{author}{\bibfnamefont{B.}~\bibnamefont{Kleine}},
  \bibinfo{author}{\bibfnamefont{G.~S.} \bibnamefont{Uhrig}}, \bibnamefont{and}
  \bibinfo{author}{\bibfnamefont{E.}~\bibnamefont{M\"uller-Hartmann}},
  \bibinfo{journal}{Europhys. Lett.} \textbf{\bibinfo{volume}{31}},
  \bibinfo{pages}{37} (\bibinfo{year}{1995}).

\bibitem[{\citenamefont{Chen and Jayaprakash}(1998)}]{chen98}
\bibinfo{author}{\bibfnamefont{K.}~\bibnamefont{Chen}} \bibnamefont{and}
  \bibinfo{author}{\bibfnamefont{C.}~\bibnamefont{Jayaprakash}},
  \bibinfo{journal}{Phys. Rev. B} \textbf{\bibinfo{volume}{57}},
  \bibinfo{pages}{5225} (\bibinfo{year}{1998}).

\bibitem[{\citenamefont{Bulla et~al.}(2008)\citenamefont{Bulla, Costi, and
  Pruschke}}]{bulla08}
\bibinfo{author}{\bibfnamefont{R.}~\bibnamefont{Bulla}},
  \bibinfo{author}{\bibfnamefont{T.~A.} \bibnamefont{Costi}}, \bibnamefont{and}
  \bibinfo{author}{\bibfnamefont{T.}~\bibnamefont{Pruschke}},
  \bibinfo{journal}{Rev. Mod. Phys.} \textbf{\bibinfo{volume}{80}},
  \bibinfo{pages}{395} (\bibinfo{year}{2008}).

\bibitem[{\citenamefont{Kehrein}(1998)}]{kehre98b}
\bibinfo{author}{\bibfnamefont{S.~K.} \bibnamefont{Kehrein}},
  \bibinfo{journal}{Phys. Rev. Lett.} \textbf{\bibinfo{volume}{81}},
  \bibinfo{pages}{3912} (\bibinfo{year}{1998}).

\bibitem[{\citenamefont{Schmidt and Uhrig}(2003)}]{schmi03c}
\bibinfo{author}{\bibfnamefont{K.~P.} \bibnamefont{Schmidt}} \bibnamefont{and}
  \bibinfo{author}{\bibfnamefont{G.~S.} \bibnamefont{Uhrig}},
  \bibinfo{journal}{Phys. Rev. Lett.} \textbf{\bibinfo{volume}{90}},
  \bibinfo{pages}{227204} (\bibinfo{year}{2003}).

\bibitem[{\citenamefont{Bl\"umer and Kalinowski}(2005)}]{blume05a}
\bibinfo{author}{\bibfnamefont{N.}~\bibnamefont{Bl\"umer}} \bibnamefont{and}
  \bibinfo{author}{\bibfnamefont{E.}~\bibnamefont{Kalinowski}},
  \bibinfo{journal}{Phys. Rev. B} \textbf{\bibinfo{volume}{71}},
  \bibinfo{pages}{195102} (\bibinfo{year}{2005}).

\bibitem[{\citenamefont{Moeller et~al.}(1995)\citenamefont{Moeller, Si,
  Kotliar, Rozenberg, and Fisher}}]{moell95}
\bibinfo{author}{\bibfnamefont{G.}~\bibnamefont{Moeller}},
  \bibinfo{author}{\bibfnamefont{Q.}~\bibnamefont{Si}},
  \bibinfo{author}{\bibfnamefont{G.}~\bibnamefont{Kotliar}},
  \bibinfo{author}{\bibfnamefont{M.}~\bibnamefont{Rozenberg}},
  \bibnamefont{and} \bibinfo{author}{\bibfnamefont{D.~S.}
  \bibnamefont{Fisher}}, \bibinfo{journal}{Phys. Rev. Lett.}
  \textbf{\bibinfo{volume}{74}}, \bibinfo{pages}{2082} (\bibinfo{year}{1995}).

\bibitem[{\citenamefont{Noack and Gebhard}(1999)}]{noack99}
\bibinfo{author}{\bibfnamefont{R.~M.} \bibnamefont{Noack}} \bibnamefont{and}
  \bibinfo{author}{\bibfnamefont{F.}~\bibnamefont{Gebhard}},
  \bibinfo{journal}{Phys. Rev. Lett.} \textbf{\bibinfo{volume}{82}},
  \bibinfo{pages}{1915} (\bibinfo{year}{1999}).

\end{thebibliography}
\end{document}